\documentclass[sigconf,twocolumn]{acmart}
\usepackage[utf8]{inputenc}
\usepackage{booktabs} 
\usepackage{tikz-uml}
\usepackage{tikz}
\usetikzlibrary{positioning}
\usepackage[htt]{hyphenat}
\usepackage{graphicx}
\usepackage{multirow}
\usepackage{algorithm}
\usepackage[noend]{algpseudocode}
\usepackage{makecell}

\DeclareUnicodeCharacter{207A}{⁺}
\newcommand{\apply}{\ensuremath{\odot}}
\tikzumlset{fill class=white}
\tikzumlset{fill package=white}
\tikzumlset{font=\footnotesize\rmfamily}

\usepackage{xcolor}  

\title{CobbleDB: Modelling Levelled Storage by Composition}

\author{Emilie Ma}
\orcid{0009-0005-3322-0805}
\affiliation{%
    \institution{University of British Columbia}
    \city{Vancouver}
    \country{Canada}
}
\email{contact@emilie.ma}

\author{Ayush Pandey}
\orcid{0000-0001-8011-5411}
\affiliation{%
    \institution{Télécom SudParis}
    \city{Palaiseau}
    \country{France}
}
\email{ayush.pandey@telecom-sudparis.eu}

\author{Annette Bieniusa}
 \orcid{0000-0002-1654-6118}
 \affiliation{
   \institution{RPTU University of Kaiserslautern-Landau}
   \city{Kaiserslautern}
   \country{Germany}
   }
 \email{bieniusa@cs.uni-kl.de}

\author{Marc Shapiro}
\orcid{0000-0002-8953-9322}
\affiliation{%
    \institution{Sorbonne-Université---LIP6 \& Inria}
    \city{Paris}
    \country{France}
}
\email{marc.shapiro@acm.org}

\ccsdesc[500]{Information systems~Storage architectures}
\ccsdesc[300]{General and reference~Verification}
\ccsdesc[300]{General and reference~Design}
\ccsdesc[300]{Theory of computation~Database theory}

\begin{document}

\acmYear{2026}\copyrightyear{2026}
\setcopyright{cc}
\setcctype[4.0]{by}
\acmConference[PaPoC '26]{Workshop on Principles and Practice of Consistency for Distributed Data}{April 27--30, 2026}{Edinburgh, Scotland, UK}
\acmBooktitle{Workshop on Principles and Practice of Consistency for Distributed Data (PaPoC '26), April 27--30, 2026, Edinburgh, Scotland, UK}
\acmDOI{10.1145/3806077.3806696}
\acmISBN{979-8-4007-2637-8/26/04}

\begin{abstract}
  We present a composition-based approach to building correct-by-construction database backing stores.
  In previous work, we specified the behaviour of several store variants and proved their correctness and equivalence.
  Here, we derive a Java implementation: the simplicity of the specification makes manual construction straightforward.
  We leverage spec-guaranteed store equivalence to compose performance features, then demonstrate practical value with CobbleDB, a reimplementation of RocksDB's levelled storage.
\end{abstract}

\maketitle

\section{Introduction}
\label{sec:introduction}

Rigorous development techniques, based on formal specifications, increase confidence that
code is bug-free, particularly in critical applications like databases.
However, verifying systems from scratch is
daunting: RocksDB's core backing store is 
$\approx\!\!300$\,KLoC \cite{rocksdb_2025}, and Redis's core is $\approx\!\!100$\,KLoC \cite{redis_2025}. 

This paper builds upon our prior formal specification framework for
database backing stores and a generic transactional protocol \cite{formel:db:sh234}.
We proved that this specification avoids anomalies, and that store lookup returns the formally-specified value.
Our first contribution in this paper is a Java implementation, derived by hand 
directly from this spec (\S~\ref{sec:implementing-stores}).
Our specification uses a common store API that encapsulates specific store variants.
This enables us to compose basic stores into high-level features, such as
write-ahead logging or data compaction.
Thus, we can improve store performance
while retaining the same correctness guarantees (\S~\ref{composing-stores}).

Our second contribution in this paper is CobbleDB, a basic
reimplementation of RocksDB's levelled storage, under this formal
framework (\S~\ref{rocksdb-emulation}).
CobbleDB supports levelled persistent storage and compaction, guarantees
Transactional Causal Consistency (TCC) or Snapshot Isolation (SI)
\cite{formel:db:rep:1856}, and achieves reasonable (for Java)
performance.
CobbleDB demonstrates how naturally our formal framework describes
existing system models.

Our generic, uniform spec makes it simple to compose the optimizations that are essential for a practical database backend.
Composition provides an approachable onramp to implement complex features with provable rigor.

\section{Formal Background}

\label{sec:formal-background}

This section summarises our formal model \cite{formel:db:sh234}.
We abstract an update as an \emph{effect} $\delta$, a function from a key's pre-value to its post-value.
There are assignment effects, like $\delta_{\text{assign 5}}$,
as well as increment effects, such as $\delta_{\text{incr 10}}$.
Using increment effects to perform updates blind (i.e. regardless of the original value) avoids the expensive
read-modify-write cycle required to implement increments via assignments.

We specify both the sequential and the concurrent behaviour of effects.
Effects combine sequentially with the associative operator $\apply$
(pronounced apply).
For instance, $\delta_{\text{assign 2}} \apply \delta_{\text{incr 1}} =
\delta_{\text{assign 3}}$.
Concurrent effects are \emph{merged}: $\mathsf{merge}(\delta_{\text{incr
    1}}, \delta_{\text{incr 3}}) = \delta_{\text{incr 4}}$.
To ensure that concurrent updates lead to a deterministic final state, a \textsf{merge} operation is associative, commutative, and idempotent, as in Conflict-Free Replicated Datatypes \cite{syn:rep:sh143}.

We formalise a transactional semantics, driven by client requests, and
guaranteeing TCC \cite{formel:db:rep:1856}.
A transaction begins; reads from a \emph{snapshot timestamp}
into a local buffer; applies effects, which it buffers locally; and
terminates, either by aborting, or by making its buffered effects
visible with a \emph{commit timestamp}.
Each action in the transaction semantics calls into the corresponding
method of the \emph{store} class, i.e., \textsf{doBegin, doUpdate,
  doAbort, doCommit}. The \textsf{Lookup} method loads the
snapshot.
A snapshot includes all effects with commit timestamps strictly less
than the snapshot's timestamp. We capture this \emph{visibility}
relationship as a DAG, called a \emph{trace}.
Commit timestamps are unique, and a commit timestamp may not be less than
any already-started snapshot (the \textsf{noInversion} constraint)
\cite{db:syn:1878}.

The \emph{valuation} function specifies the expected value of a key at some point in the trace.
Starting from an empty state, the sequential (resp.~concurrent) effects
visible at that point are \textsf{apply}'ed (resp.~\textsf{merge}'d).
Importantly, any effect before the most recent assignment can be ignored.

\subsection{Specifying Backing Stores}
\label{sec:specifying-stores}

A backing store is the database component that reliably stores and
retrieves data, by implementing the store methods outlined above.
Our specification formalises the behaviour of classical \emph{map-based}
and \emph{journal-based} backing stores.
Both maintain a partially ordered set of effects.

The \emph{map store} variant is designed for read-heavy workloads.
It maps a key to a list of effects, with metadata to distinguish concurrent effects.
Uncommitted writes remain transaction-private, i.e., \textsf{doUpdate} is a no-op; \textsf{doCommit} pushes all its buffered effects in bulk.
\textsf{Map.Lookup} searches for the most recent assignment(s) up to a specified snapshot timestamp, and \textsf{applies}{\slash}\textsf{merge}s increments that follow.

The \emph{journal store} variant is a sequential structure ordered by time; it is optimised for writes and is crash tolerant.
A \textsf{doX} method (e.g., \textsf{doUpdate}) appends the corresponding event to its log.
\textsf{Journal.Lookup} replays updates up to the specified timestamp; similarly for crash recovery.

We prove elsewhere \cite{formel:db:sh234} that
map and journal stores' \textsf{Lookup} conform to the valuation.
This implies that the map and journal specifications are correct; thus
that they behave identically; and thus that they are composable.
Formalising store composition enables representing the complex structure of modern databases, as discussed later.

\section{Implementing Basic Stores}
\label{sec:implementing-stores}

We implement the above spec in Java; the code is {open-source} \cite{cobbledb-source}.
We chose Java for its widely-used concurrency libraries and to leverage the Java-based YCSB benchmarks, discussed later.
The implementation has no optimisations other than those mentioned herein.

Because the spec governs observable behaviour, implementation correctness reduces to conformance.
For every proposition in the spec, the code has a matching \textsf{assert}.
We also implemented extensive test suites, with 100\% code coverage.
Although the implementation is derived manually, correctness can thus be systematically validated.

\paragraph{Store logic}
\label{sec:store-logic}

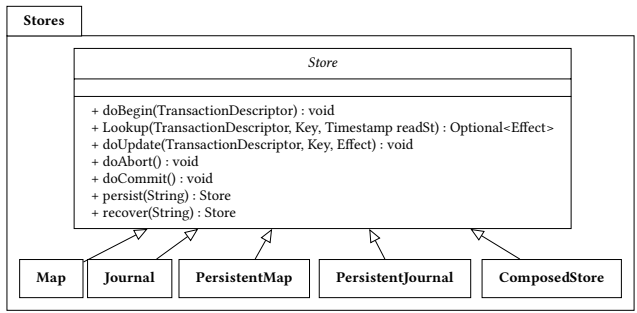
\begin{figure}
    \centering
    \resizebox{\columnwidth}{!}{
    \begin{tikzpicture}
    \begin{umlpackage}{Stores}
    \umlclass[type=abstract]{Store}{}{ 
      + doBegin(TransactionDescriptor) : void  \\
      + Lookup(TransactionDescriptor, Key, Timestamp readSt) : Optional<Effect>  \\
      + doUpdate(TransactionDescriptor, Key, Effect) : void  \\
      + doAbort() : void  \\
      + doCommit() : void  \\
      + persist(String) : Store  \\
      + recover(String) : Store  \\
    } 
    \umlsimpleclass[below left=.5cm and 4.5cm of Store.south, anchor=north, width=.8cm]{Map}
    \umlsimpleclass[below left=.5cm and 3.20cm of Store.south, anchor=north, width=.8cm]{Journal}
    \umlsimpleclass[below left=.5cm and 1.30cm of Store.south, anchor=north]{PersistentMap}
    \umlsimpleclass[below left=.5cm and -1.2cm of Store.south, anchor=north]{PersistentJournal}
    \umlsimpleclass[below left=.5cm and -3.8cm of Store.south, anchor=north]{ComposedStore}
    \umlinherit{Map}{Store}
    \umlinherit{Journal}{Store}
    \umlinherit{PersistentMap}{Store}
    \umlinherit{PersistentJournal}{Store}
    \umlinherit{ComposedStore}{Store}
    \end{umlpackage}
    \end{tikzpicture}
    }
    \caption{Basic stores class hierarchy diagram.}
    \label{fig:store-types}
\end{figure}

As shown in Figure~\ref{fig:store-types}, each concrete \emph{basic store} (e.g. \textsf{Map}, \textsf{Journal}) extends the abstract class \textsf{Store}.

The in-memory \textsf{Journal} type is implemented as a thread-safe deque of records; each \textsf{doX{}} API method appends a corresponding record.
\textsf{Journal.Lookup} extracts the committed effects, partially-orders them by visibility (based on begin and commit record timestamps), and \textsf{applies} or \textsf{merge}s them as described earlier.
The \textsf{Map} is implemented as a thread-safe hashmap of keys to partially-ordered deques of versions.
\textsf{Map.Lookup} follows the journal algorithm, applied per-key.

We implement persistent variants of the in-memory stores.
For example, the persistent journal appends records to a sequential file.
To ensure crash-atomicity, \textsf{doCommit} flushes the file.
On recovery, it checks integrity (using sentinel metadata in each
record) and appends an abort record to any unterminated transaction, then proceeds normally.
In contrast, a map is made read-only before being persisted, ensuring
it can be recovered intact.
In our usage (\S~\ref{rocksdb-emulation}), a persistent store is
rarely read and its size is bounded. Thus, its efficiency is not a main
concern. We use standard Java serialisation.

\paragraph{Transaction protocol logic}
\label{sec:transaction-logic}

\begin{figure}
    \centering
    \resizebox{\columnwidth}{!}{
    \begin{tikzpicture}
    \umlclass{TransactionDescriptor}{ 
        + String TxnId \\
        + Timestamp st // snapshot \\
        + Timestamp ct // commit \\
        + Map<Key, Effect> readBuffer \\
        + Set<Key> initSet \\
        + Map<Key, Effect> effectBuffer \\
    }{}
    \umlclass[right=0.8cm of TransactionDescriptor.north east,anchor=north west]{TransactionCoordinator}{}{ 
        + beginTxn(String, Timestamp): String \\
        + read(Key): Value \\
        + update(Key, Effect): void \\
        + abort(): void \\
        + commit(Timestamp): void
    } 
    \umlclass[below=0.6cm of TransactionCoordinator.south]{TransactionManager}{
        + Map<String, TxnCoordinator> txns \\
        \# TimestampGenerator tsGen
    }{ 
        + beginTxn(): TxnCoordinator \\
        + abortTxn(String): void \\
        + commitTxn(String): void
    } 
    \umlclass[below=0.6cm of TransactionDescriptor.south]{TimestampGenerator}{}{ 
        + next(): Timestamp \\ 
        + peek(): Timestamp \\
        + endCommit(Timestamp) \\
    } 
    \umlunicompo[mult1=1,mult2=1]{TransactionCoordinator}{TransactionDescriptor}
    \umlunicompo[mult1=1,mult2=*,arg=creates]{TransactionManager}{TransactionCoordinator}
    \umluniassoc[mult1=1,mult2=1]{TransactionManager}{TimestampGenerator}
    \end{tikzpicture}
    }
    \caption{Transaction layer class diagram.}
    \label{fig:txn-types}
\end{figure}
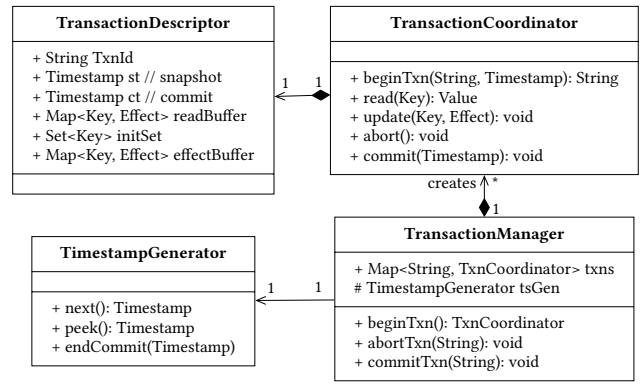

Figure~\ref{fig:txn-types} depicts transaction-related types.
A \textsf{TransactionDescriptor} tracks the state of a transaction, including its unique identifier, snapshot and commit timestamps, and read and write buffers.
The \textsf{TransactionCoordinator} (TC) directly implements the transaction spec. 
The \textsf{TransactionManager} (TM) handles the lifecycle of allocating{\slash}deallocating a TC per client.
This exemplifies the ease of development: given that the basic stores implement the same APIs and satisfy the same properties, the Java translation of the transaction protocol ``falls out'' of its spec.

A central \textsf{TimestampGenerator} ensures the timestamp constraints (see \S~\ref{sec:formal-background}).
In particular, for \textsf{noInversion}, it maintains a \emph{minimum allowed commit timestamp} always greater than the \emph{maximum allowed snapshot timestamp}.
Our logic in Algorithm~\ref{alg:tsgen} is inspired by the design considerations of \citet{hatia_2023}.
\textsf{next} issues a unique commit timestamp, while \textsf{endCommitNotify} lazily advances the snapshot timestamp after earlier timestamp requests are durably committed. 
The TC's \textsf{doCommit} is called in between these two methods: splitting up the calls keeps \textsf{next} non‑blocking and ensures crash-atomicity by finalizing commits only after a durable flush.
If a flush is delayed and transactions are pending when \textsf{endCommitNotify} is called, the max snapshot timestamp does not update until the ``gap'' is closed, yielding at most older snapshots but never a partially-flushed timestamp.

\algdef{SE}[SYNC]{SyncProcedure}{EndSyncProcedure}[2]{%
  \textbf{synchronized}\ \algorithmicprocedure\ #1(#2)}{}

\begin{algorithm}[tp]
  \small
  \caption{\texttt{TimestampGenerator} pseudocode}
\label{alg:tsgen}
\begin{algorithmic}
    \State $\mathit{minAllowedCT} \gets \text{AtomicInt(0)}$ \Comment{CT: commit timestamp}
    \State $\mathit{maxAllowedST} \gets \text{AtomicInt(0)}$ \Comment{ST: snapshot timestamp}
    \State $\mathit{running} \gets \varnothing$ \Comment{CT's of pending transactions}
    \State $\triangleright$ \texttt{beginTxn} sets snapshot to \texttt{maxAllowedST}
    \State $\triangleright$ \texttt{commitTxn} leases commit timestamp with \texttt{next}, then finalizes\\\hspace{.8em}with \texttt{endCommitNotify}
    \Procedure{next}{} 
    \State $\mathit{ts} \gets \mathit{minAllowedCT}\text{.getAndIncrement()}$
    \State $\mathit{running}$.add($\mathit{ts}$)
    \State \Return $\mathit{ts}$
    \EndProcedure
    \SyncProcedure{endCommitNotify}{$\mathit{CT}$} 
    \State $\mathit{running}$.remove($\mathit{ts} \in \mathit{running} | \mathit{ts} \leq \mathit{CT} \land \mathit{ts}$ committed)
    \State \algorithmicif\ {any ts removed} \algorithmicthen\  {$\mathit{maxAllowedST} \gets \min(\mathit{running})$}
    \EndSyncProcedure
\end{algorithmic}
\end{algorithm}

Following the formal model, CobbleDB guarantees TCC\@.
By strengthening the checks in \textsf{commitTxn}, it also supports SI\@.
Supporting PSI or SSER would be straightforward
\cite{formel:db:rep:1856}.

\section{Composing Stores}
\label{composing-stores}

Recall from \S~\ref{sec:formal-background} that basic stores are provably equivalent: passing a \textsf{Map} or a \textsf{Journal} to a \textsf{TransactionManager} yields the same behaviour.
Hence, we can compose stores to improve performance while observing identical results.
For instance, we might serve reads from an in-memory map while
persisting writes in a journal; this implements a \emph{write-ahead log}
that performs efficient sequential writes and avoids disk I\slash{}O on
read.
Composition helps specify high-level features; low-level implementation details are out of scope.

A composed store follows the same \textsf{Store} API as a basic store.
We call its component stores ``ministores.''
A ministore tracks a slice of the database history; we call
\emph{window} the closed-open interval of timestamps for which the
ministore contains all effects.
A composed store's window is the union of its ministore windows. The top-level window must cover the entire history: gaps would render \textsf{Lookup} unsafe.
A ministore is significant only in its window, and equivalence holds
where windows overlap.

Ministores follow the ``write all, read one'' rule.
Thus, a \textsf{doX} method on a composed store recursively calls the same method on \emph{all} its ministores in the window.
This maintains equivalence between ministores within overlaps.
Because the ministore windows collectively cover the whole top-level history, a composed store is thus always observationally equivalent under \textsf{Lookup} to a basic store.
Conversely, ministores with the same window are interchangeable for
\textsf{Lookup}; we prefer to read from the fastest one.
By the equivalence of store \textsf{Lookup}s, this optimization does not affect correctness.

\section{Emulating RocksDB's Levelled Storage}
\label{rocksdb-emulation}

To demonstrate composition, we implement CobbleDB, which emulates RocksDB by
composing basic stores.
RocksDB is a persistent KV store offering SI \cite{rocksdb_2025}.%
\footnote{
  The spec of \S~\ref{sec:formal-background} supports TCC; CobbleDB
  strengthens this to SI for a fair comparison with RocksDB\@.
}
 Its backing store is based on a Log-Structured Merge-Tree.
Its \emph{levels} contain updates from most recent (at the top) to oldest (at the bottom).
Each level contains a sequence of files.
The top-most \emph{live level} receives events in its rightmost file; the remaining live level files and all lower levels (numbered from 0) are immutable except via compaction.
RocksDB uses levelled compaction: from the live level into L0, from L0 into L1, and so on \cite{rocksdb_compaction_2025}.
Compaction is instrumental for RocksDB's performance.

CobbleDB implements levelled storage by composing stores.
This is justified by the equivalence proof and made easy by the uniform API.
CobbleDB does not implement RocksDB's other optimisations (e.g., indexing, Bloom filters).

\paragraph{Implementation}
\label{sec:implementation}

CobbleDB relies on two additional store types: a \textsf{Checkpoint} and a \textsf{WALMemtablePair}.
A \textsf{Checkpoint} is a special map with a single version per key, recording its value at checkpoint time.
A basic store can be compacted into a \textsf{Checkpoint} via its \textsf{checkpoint()} method, which returns \textsf{Lookup} on each key at the high timestamp of its window.

A \textsf{WALMemtablePair} (WMP) is an example of composition, combining a persistent journal with an in-memory map.%
\footnote{
  WAL and Memtable are RocksDB's names for our Persistent Journal and In-Memory Map, respectively.
}
\textsf{WMP}s make up the topmost, \emph{live level} of the backing store.
A \textsf{doUpdate} appends the update to the journal only (recall that \textsf{Map.doUpdate} is a no-op), whereas \textsf{doCommit} also pushes all the transaction's updates to the map.
On the other hand, \textsf{Lookup} is routed to the map for fast reads.

See Algorithm~\ref{alg:lookup} for CobbleDB's lookup pseudocode.%
\footnote{
  The use of composition and the data model come from the formal
  specification.
  The specific algorithms are modelled after those of RocksDB.
}
Lookup proceeds from most to least recent update.
It reads the \textsf{WMPs} in the live level,
accumulating effects and stopping at the first assignment.
If no assignment is reached in the live level, it proceeds similarly
through L0, then to L1, and so on.
The store finally \textsf{applies} the effects in sequence.

Our implementation restricts concurrency to ensure that every
transaction starts and finishes within the same ministore.
Then, there may be concurrent transactions to be \textsf{merge}'d within
the same ministore only, and no concurrency between ministores or
levels.
As a result, we can consolidate effects per ministore and simply
\textsf{apply} these in sequence.
\begin{algorithm}[tp]
\small
\caption{CobbleDB's levelled lookup algorithm}
\label{alg:lookup}
\begin{algorithmic}
    \Procedure{Lookup}{$key, readSt$} 
    \State $\sigma \gets \text{CobbleDB Store}$
    \State $\delta\text{[]} \gets \text{[]}$ \Comment{list of effects}
    \State $\mathit{level} \gets -1$ \Comment{start at the live level}
    \While{no assignment in $\delta\text{[]}$ and $\mathit{level} < \mathit{MAX}$}
        \State $\sigma_{\mathit{level}} := \sigma[\mathit{level}]$ \Comment{list of ministores at level}
        \For{$n := len(\sigma_{\mathit{level}}) - 1; n \geq 0; n--$}
            \State $\delta_r \gets \sigma_{\mathit{level}}$[$n$].Lookup(key, readSt)
            \If{$\delta_r \neq \bot$ \Comment{if an effect is found}} 
                \State $\delta$[].push($\delta_r$)
                \State \algorithmicif {$\delta_r.isAssignment()$} \algorithmicthen\  {\textbf{break}}
            \EndIf
        \EndFor
        \State $\mathit{level} \gets \mathit{level} + 1$
    \EndWhile
    \State \Return $\delta[n] \apply \dots \apply \delta[0]$
    \EndProcedure
\end{algorithmic}
\end{algorithm}

\paragraph{Compaction}
\label{sec:compaction}

RocksDB's levelled compaction reduces read amplification and storage requirements.
Each level has a storage capacity, which triggers compaction when it is exceeded \cite{rocksdb_compaction_2025}.
L0 contains newly flushed data from the live level, sharded by timestamp; other levels are sharded by key.

See Algorithm~\ref{alg:compaction} for CobbleDB's compaction pseudocode.
At the live level, compaction checkpoints the oldest stores until the remaining stores fit within the live level's capacity.
The checkpoints are pushed to L0 and the source stores are removed.
If this causes an overflow at L0, compaction continues into lower levels until the stores at each level fit within capacity.
For each checkpoint on level $N$, compaction finds the corresponding checkpoint on level $N+1$ and \textsf{applies} updates in sequence.
No concurrent transactions exist across stores or levels, so compaction does not require effect \textsf{merge}s. 


\begin{algorithm}[tp]
\small
\caption{CobbleDB's compaction algorithm}
\label{alg:compaction}
\begin{algorithmic}
    \Procedure{compact}{}
    \State $\sigma \gets \text{CobbleDB Store}$
    \State $\sigma_{-1}[] \gets \text{list of overflowing stores from start of live level}$
    \State $\sigma_{-1}[]$.map($s \rightarrow s$.checkpoint())
    \State $\sigma_{0}[]$.push(\ldots $\sigma_{-1}[]$), remove $\sigma_{-1}[]$ from live level
    \State $\mathit{level} \gets 0$ \Comment{continue compaction at level 0}
    \While{$\mathit{level} < \mathit{MAX} - 1$}
        \State $\sigma_{\mathit{level}}[] \gets \text{list of overflowing stores from level $\mathit{level}$}$
        \For{each store $s \in \sigma_{\mathit{level}}[]$}
            \For{each store $s_{\mathit{low}} \in \sigma_{\mathit{level + 1}}[]$}
                \For{each key $k$ in $s$}
                    \If{$s_{\mathit{low}}$ contains $k$} 
                        \State $s_{\mathit{low}}[k] \gets s_{\mathit{low}}[k] \apply s[k]$
                    \EndIf
                \EndFor
            \EndFor
            \State create $s_{\mathit{new}}$ to store any keys not yet compacted
            \State $\sigma_{\mathit{level + 1}}[]$.push($s_{\mathit{new}}$)
            \State remove $\sigma_{\mathit{level}}[]$ from level $\mathit{level}$
        \EndFor
        \State $\mathit{level} \gets \mathit{level} + 1$
    \EndWhile
    \EndProcedure
\end{algorithmic}
\end{algorithm}

\paragraph{Recovery}
\label{sec:recovery}

RocksDB records its internal file structure in an authoritative
\emph{MANIFEST log} \cite{rocksdb_manifest_2025}.
Similarly, CobbleDB's MANIFEST, a persistent journal, records ministore
locations in a crash-tolerant manner.
On crash recovery, the MANIFEST journal retrieves each level's file
paths.
Each store is recovered, and the live level is checkpointed directly to L0, matching RocksDB's behaviour.
Recovery requires updating the MANIFEST with the new L0 checkpoints' paths, as well as clearing the live level's old paths.
The snapshot and commit timestamp of this MANIFEST transaction are equal to the highest timestamp encountered during recovery.
The \textsf{TransactionManager} then sets its timestamp generator to only pick timestamps after this timestamp.

Our recovery implementation is idempotent, thus tolerating repeated crash-recovery failures.
For instance, if serialization fails or the system crashes while writing a ministore file, the partially written file is not referenced by a committed MANIFEST transaction and is ignored.
Similarly, if disk writes fail while checkpointing the live level; the new L0 checkpoints will be re-created on future recovery attempts.
Our tests for persistence and recovery include serialization and disk I\slash{}O failure simulations.

\section{Results}
\label{sec:results}

Our uniform, composable specification allows us to naturally express RocksDB's levelled storage structure.
CobbleDB comprises 3\,204\,LoC in Java\footnote{In comparison, RocksDB's
  core is $\approx\!\!300\,000$\,LoC of C++, even excluding, to the best
  of our ability, features that CobbleDB does not implement (e.g., compression). While this is not an apples-to-apples comparison, this still demonstrates our spec-driven approach's brevity. We exclude tests and libraries from LoC in both cases.}.
We apply the most straightforward data structures without concern for performance or optimization; for instance, a timestamp is represented as a map.
Our tests have 100\% code coverage, exercising both sequential behaviour and concurrent execution with up to ten workers. 
This includes scenarios with concurrent reads, writes, and commits and stress tests for composition consistency.
While computing concurrent behaviour coverage is infeasible, our test suite covers the actions most critical for correctness.

The following sections present performance evaluations:
\begin{itemize}
    \item \textbf{RQ1}: How does CobbleDB compare to basic stores?
    \item \textbf{RQ2}: How does CobbleDB compare to RocksDB?
    \item \textbf{RQ3}: How do CobbleDB's configurations compare?
\end{itemize}

\paragraph{RQ1: CobbleDB vs.~basic backing stores}

We benchmark stores by extending the YCSB framework \cite{YCSB_2025} with custom transactional workloads.
We run each workload with 1--20 YCSB threads, each communicating with its own database client over HTTP (\texttt{TCP\_NODELAY} enabled).
A transaction contains five lookup or update operations.
The key distribution is zipfian.
Benchmarks are run on machines with two Intel Xeon E5-2620 v4 CPUs and 64\,GB RAM on a 10\,Gbps network.
YCSB and the store run on separate machines to avoid noise; given network latency and older hardware, our results will differ from published benchmarks using a single machine.
Each experiment starts with an empty database (no warmup) and runs for 60\,s.
We define three workloads:
\begin{itemize}
    \item \textsf{txn}: 50\slash{}50 split between lookups and assignments.
    \item \textsf{old\_reads}: same, with a 50\% chance that the snapshot timestamp is strictly before the last commit.
    \item \textsf{txn\_increments}: 50\slash{}30\slash{}20 split between lookups, assignments, and increments respectively.
\end{itemize}

\begin{figure}[t]
    \centering
    \includegraphics[width=\columnwidth]{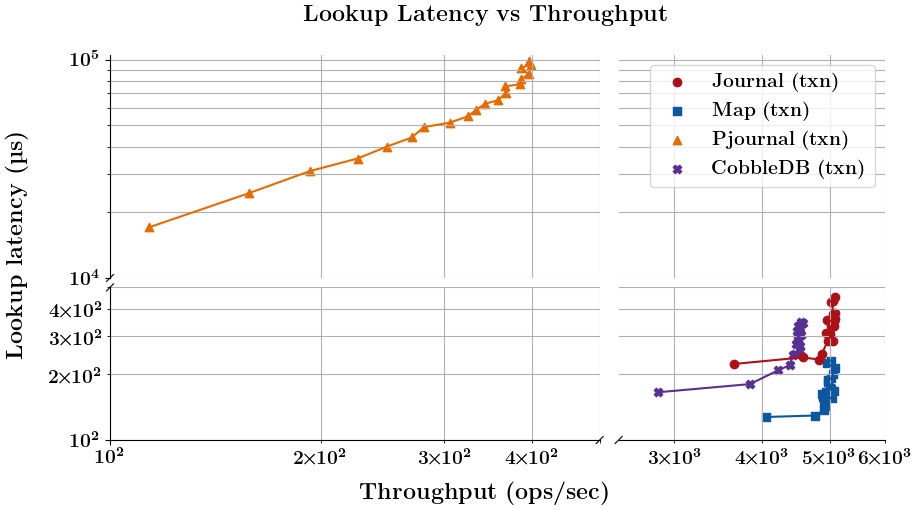}
    \includegraphics[width=\columnwidth]{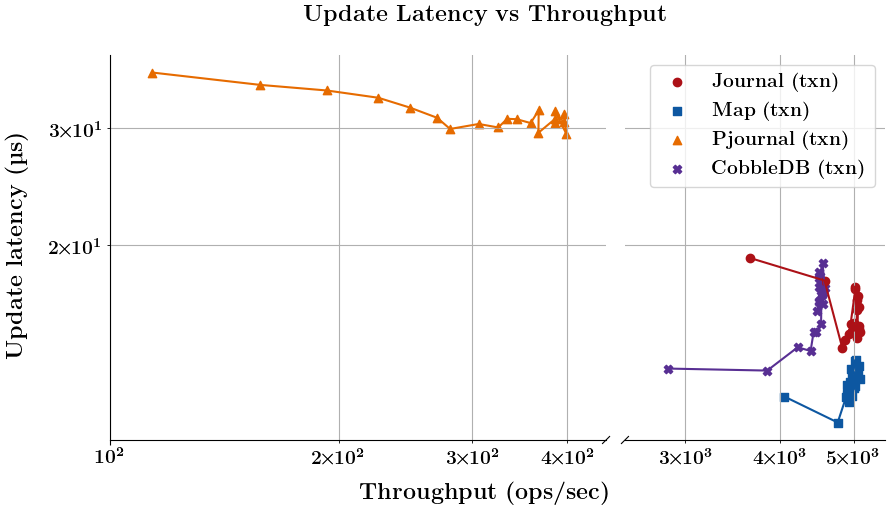}
    \caption{Latency vs.~throughput for CobbleDB and basic stores.}
    \label{fig:tl-basic}
\end{figure}

Figure~\ref{fig:tl-basic} shows latency and throughput for CobbleDB and the basic stores, under the \textsf{txn} workload.\footnote{We omit the persistent map, as it is not used in the CobbleDB composition. We also omit ranges without data points via axis breaks for visual clarity.}
Each point represents a separate run; the number of YCSB client threads varies from 1~to 20.
The journal, map, and CobbleDB have similar performance trends, with a latency-throughput knee at 2--4 threads.
The map has lower latency than the journal: its key-sharded structure leads to smaller effect lists to process.
Otherwise, the two have comparable max throughputs of 5\,088\,ops\slash{}s (journal) and 5\,098\,ops\slash{}s (map).
CobbleDB's lookup latency is lower than the journal's: checkpoint compaction decreases the reads required per lookup.
Compaction overhead also leads to lower throughput.

The persistent journal's performance is orders of magnitude worse. Lookup latency increases with additional threads due to extra log records, and flush-on-commit caps throughput.
The persistent journal also has worse update latency than CobbleDB\@.
This is surprising, since CobbleDB first writes to the persistent
journal component of \textsf{WMP}.
However, when increasing writes to 100\% (experiment not plotted), update latency is nearly identical.
We surmise that the JVM serialises file access.
While CobbleDB serves reads from \textsf{WMP} map separately, in the
persistent journal writes must wait for reads.
CobbleDB's max throughput is 4\,581\,ops\slash{}s whereas the persistent journal's is 398\,ops\slash{}s, providing $11.5\times$ more throughput with the same persistence guarantees.

\paragraph{RQ2: CobbleDB vs.~RocksDB}

We ran the same YCSB workloads to compare CobbleDB against RocksDB in its default configuration. We note that the results are not directly comparable. For one, RocksDB is implemented in optimized C++ instead of Java.
The latest Computer Language Benchmark Game (CLBG) finds the most optimized Java programs to be $\approx\!\!4.5\times$ slower than C++ for high I\slash{}O workloads (up to 12x for lower I\slash{}O workloads) \cite{clbg}.
As well, CobbleDB lacks many of RocksDB's performance features and was deliberately designed to reflect our spec without optimizations. 

\begin{figure}[t]
    \centering
    \includegraphics[width=\columnwidth]{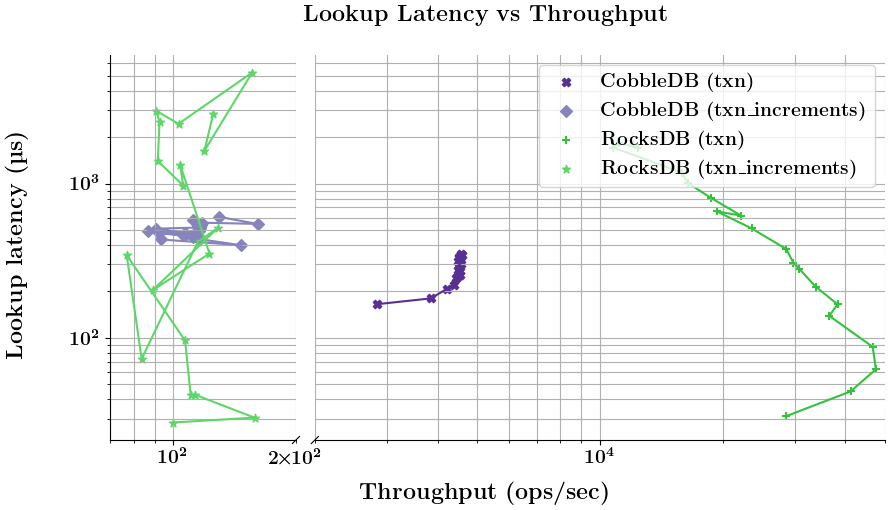}
    \includegraphics[width=\columnwidth]{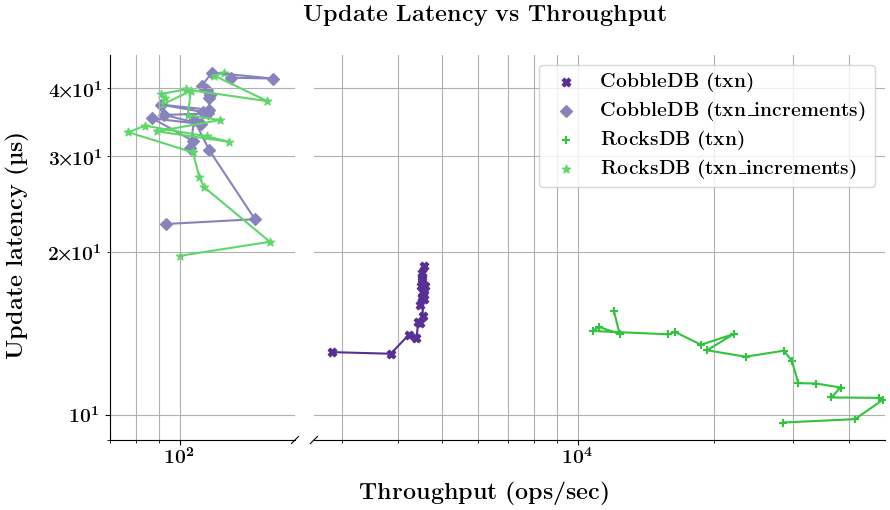}
    \caption{Latency vs. throughput for CobbleDB and RocksDB.}
    \label{fig:tl-cobble-rocks}
\end{figure}

Figure~\ref{fig:tl-cobble-rocks} compares CobbleDB and RocksDB on the \textsf{txn} and \textsf{txn\_increments} workloads.\footnote{\textsf{old\_reads} behaves similarly to \textsf{txn} and is discussed later.}
Again, each point represents a separate run, varying the number of threads from 1~to 20.

On the \textsf{txn} workload, CobbleDB's throughput is one order of
magnitude less than RocksDB's (RocksDB max throughput
47\,498\,ops\slash{}s vs.~CobbleDB 4\,581\,ops\slash{}s).
However, RocksDB's throughput decreases as thread count increases,
likely due to its write backpressure \cite{rocksdb_write_stalls_2025}.
RocksDB's lookup latencies grow faster than CobbleDB's, ballooning $59\times$ from 1~to 20~threads, compared to CobbleDB's $2\times$ increase; RocksDB's update latency grew $1.6\times$ compared to CobbleDB's $1.3\times$.

On the \textsf{txn\_increments} workload, throughput is much lower.
As shown by the spikes, latency is highly variable: merging increments accumulates more state.
The zipfian distribution used to select keys also introduces significant spread in how far back the search must progress.
Yet, CobbleDB handles increments more efficiently than RocksDB: CobbleDB's lookup latency averages $499\,\mu{}$s, compared to RocksDB's $1\,174\,\mu{}$s, and update latency matches RocksDB's.
CobbleDB's max throughput on this workload also slightly exceeds RocksDB's: 161\,ops\slash{}s v.s. RocksDB's 156\,ops\slash{}s.

In summary, RocksDB betters CobbleDB on
assignment-only workloads, but they are comparable when
increments are included.
On assignment workloads, the performance ratio is of the
order observed in the CLBG, despite CobbleDB's
spec-driven, unoptimised implementation.

\begin{figure}[t]
    \centering
    \includegraphics[width=\columnwidth]{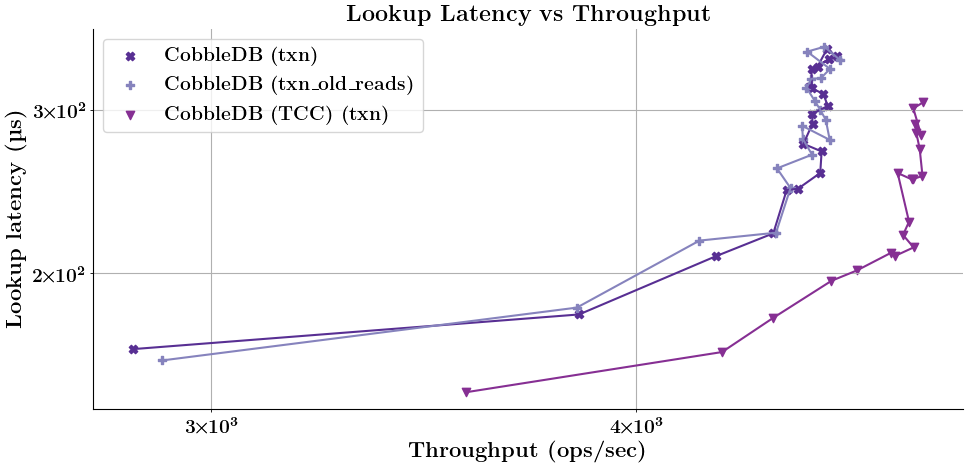}
    \includegraphics[width=\columnwidth]{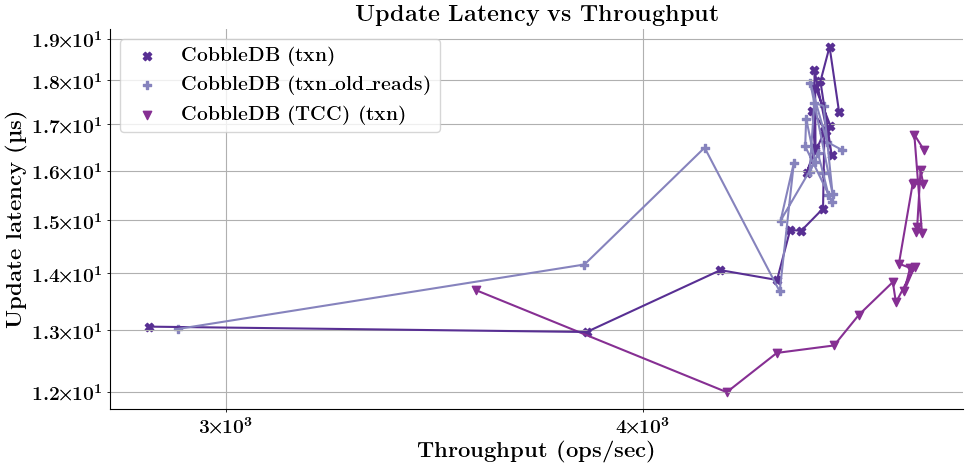}
    \caption{Latency vs.~throughput for different CobbleDB configurations and workloads.}
    \label{fig:tl-cobble-comp}
\end{figure}

\paragraph{RQ3: CobbleDB configurations}
\balance
Figure~\ref{fig:tl-cobble-comp} compares default CobbleDB (ensuring SI) against the \textsf{txn} and \textsf{old\_reads} workloads.\footnote{We omit the \textsf{txn\_increments} workload for visual clarity given significantly lower throughput.}
Thread count again varies from 1~to 20 for each workload run.
We also run CobbleDB under TCC against the \textsf{txn} workload, to exercise the concurrent updates' \textsf{merge}.

CobbleDB achieves a max throughput of 4\,592\,ops\slash{}s on the \textsf{old\_reads} workload, relative to 4\,581\,ops\slash{}s on the \textsf{txn} workload.
Despite the \textsf{old\_reads} workload requiring deeper reads, the performance trends are similar, highlighting how composition via the checkpoint stores aids performance.

Under TCC, CobbleDB exhibits similar latency-throughput curves as under SI, with performance improvements due to an increase in successful concurrent transactions.
CobbleDB's throughput peaked at 4\,859\,ops\slash{}s.
CobbleDB's update latency with TCC averages $14.4\,\mu{}$s compared to $20.8\,\mu{}$s with SI, as TCC does not check nor abort conflicting updates.
Lookup latency with TCC also decreased ($236\,\mu{}$s vs.~$328\,\mu{}$s on SI): increased updates make it more likely that lookup terminates earlier in the live, or more recent, levels.

\section{Related Work}
\label{sec:related-work}

Prior research has also verified components of new database backing stores. For example, Malecha et al.\ develop a Rocq-verified relational database \cite{formel:db:1859}, and Patel et al.\ present a framework to verify the linearizability of concurrent key-value stores \cite{syn:db:formel:1883}. However, these works do not address crash recovery or transactional guarantees, making them inadequate for modern production use. The VerIso project addresses the transactional specification gap, proposing a Isabelle\slash{}HOL framework for concurrency control protocols \cite{db:syn:1884}. It also leaves durability to future work; in contrast, CobbleDB and our spec deliberately handle these concerns.


\section{Conclusion}
\label{sec:conclusion}

In this paper, we show that composition is a promising way to improve the performance of formally-specified stores.
Because our spec guarantees the equivalence of stores in their overlapping window, composition unlocks improvements like serving reads from faster stores while retaining the same correctness assurances.
As an example, we describe CobbleDB, an emulation of RocksDB's levelled storage structure.
Our test suite, based on the spec propositions and with full code coverage, lends confidence in this implementation.
We found that CobbleDB achieved 9.6\% of RocksDB's performance, with the discrepancy likely attributable to implementation language differences.
Our implementation highlights how cleanly composition maps onto production database components, like RocksDB's levelled compaction.

Our approach is not intended to replace database development techniques wholesale, but to integrate with existing ones.
Developers can continue to hand-write systems and apply conventional testing techniques, but ground their high-level architecture in a formal spec and incrementally refactor.
For instance, they could start by refactoring to our WAL and recovery system to gain consistency guarantees, while keeping code paths for other features separate.

Future work includes emulating additional databases to further validate our approach, particularly distributed databases like AntidoteDB \cite{antidote_2025, rep:pro:sh182}. As well, we plan to continue performance tuning for CobbleDB.

By starting from a generic specification and composing basic backing stores, we can cleanly derive advanced performance features. Our approach supports the practical application of formal methods, using composition to make rigorous guarantees approachable in complex storage architectures.

\begin{acks}
  Emilie Ma conducted this work as a visiting intern at Sorbonne
  Université, based on the previous contributions of Saalik Hatia's PhD
  and of Ayush Pandey.
  We thank Jaurel Fosset for his contributions to optimizations and
  benchmarking.
  We also thank Carla Ferreira for her feedback.
  This work was carried out thanks to a generous
  \grantsponsor{amazon}{Amazon Research
    Award}{https://www.amazon.science/research-Amazon},
  and was supported in part by
  \grantsponsor{centeanes}{ANR}{https://anr.fr/} under the Centeanes grant
  \grantnum[https://anr.fr/Project-ANR-24-CE25-5598]{centeanes}{ANR-24-CE25-5598}.
  Our experiments were carried out on the Grid'5000 testbed, supported by
  a scientific interest group hosted by Inria and including CNRS, RENATER
  and several universities as well as other organizations (see
  \url{grid5000.fr}).
\end{acks}
\bibliographystyle{ACM-Reference-Format}
\bibliography{predef,references,shapiro-bib-ext,shapiro-bib-perso}

\end{document}